\documentclass[conference]{IEEEtran}
\IEEEoverridecommandlockouts
\usepackage{cite}
\usepackage{amsmath,amssymb,amsfonts}
\usepackage{algorithmic}
\usepackage{graphicx, epsfig}
\usepackage{textcomp}
\usepackage{xcolor}
\def\BibTeX{{\rm B\kern-.05em{\sc i\kern-.025em b}\kern-.08em
    T\kern-.1667em\lower.7ex\hbox{E}\kern-.125emX}}
\begin{document}

\title{Quantum Machine Learning: An Interplay Between Quantum Computing and Machine Learning\thanks{The views expressed in this article are those of the authors and do not represent the views of Wells Fargo. This article is for informational purposes only. Nothing contained in this article should be construed as investment advice. Wells Fargo makes no express or implied warranties and disclaims all legal, tax, and accounting implications related to this article.}		\\
}

\author{\IEEEauthorblockN{1\textsuperscript{st} Jun Qi}
\IEEEauthorblockA{
\textit{Hong Kong Baptist Univeristy}\\
Hong Kong, China	 \\
jun-qi@comp.hkbu.edu.hk}
\and
\IEEEauthorblockN{2\textsuperscript{nd} Chao-Han Huck Yang}
\IEEEauthorblockA{
\textit{NVIDIA Research}\\
Taipei, Taiwan \\
hucky@nvidia.com}
\and
\IEEEauthorblockN{3\textsuperscript{rd} Samuel Yen-Chi Chen}
\IEEEauthorblockA{
\textit{Wells Fargo Bank}\\
New York, NY, USA \\
yen-chi.chen@wellsfargo.com} 
\and
\IEEEauthorblockN{4\textsuperscript{th} Pin-Yu Chen}
\IEEEauthorblockA{
\textit{IBM Research}\\
Yorktown Heights, NY, USA \\
pin-yu.chen@ibm.com} 
}

\maketitle

\begin{abstract}
Quantum machine learning (QML) is a rapidly growing field that combines quantum computing principles with traditional machine learning. It seeks to revolutionize machine learning by harnessing the unique capabilities of quantum mechanics and employs machine learning techniques to advance quantum computing research. This paper introduces quantum computing for the machine learning paradigm, where variational quantum circuits (VQC) are used to develop QML architectures on noisy intermediate-scale quantum (NISQ) devices. We discuss machine learning for the quantum computing paradigm, showcasing our recent theoretical and empirical findings. In particular, we delve into future directions for studying QML, exploring the potential industrial impacts of QML research. 
\end{abstract}


\section{Introduction}

Despite their remarkable natural language processing~\cite{deng2018deep} and computer vision~\cite{szeliski2022computer} achievements, deep neural networks face computational bottlenecks for emerging applications like drug discovery~\cite{smalley2017ai} and materials science. The massive size of large language models further exacerbates this issue. As machine learning continues to advance, the limitations of classical computing are becoming more apparent, hindering progress in these fields. A promising solution on the horizon is quantum computing~\cite{nielsen2010quantum}. The advent of quantum computing holds great potential for revolutionizing or enhancing the computational efficiency of machine learning algorithms. Although the deployment of quantum computers is still in its early stage, cloud-based services provide accessible quantum computing environments, such as IBM Quantum Experience~\cite{harper2019fault} and CUDA Quantum~\cite{kim2023cuda}, enabling users to develop quantum algorithms for machine learning. 

Quantum machine learning (QML) is an interdisciplinary field that combines quantum mechanics and machine learning~\cite{biamonte2017quantum, power_data, schuld2015introduction}. Leveraging QML theories and algorithms can enhance the computational efficiency of machine learning models~\cite{liu2021rigorous}. The noisy intermediate-scale quantum (NISQ) era, with its limited number of qubits and high noise levels, presents challenges but also opportunities for exploring the potential of quantum computing for machine learning~\cite{liu2024towards, Preskill2018quantumcomputingin, cerezo2022challenges}. In particular, variational quantum circuits (VQC) constitute a QML architecture~\cite{cerezo2021variational}, such as quantum convolutional neural networks~\cite{cong2019quantum} and quantum graph neural networks~\cite{verdon2019quantum}, for data processing and making predictions. The parametric quantum circuits (PQC) in VQC can be adjustable in the training process using optimization methods like stochastic gradient descent (SGD) to minimize the cost function in a back-propagation manner. The VQC has been demonstrated to be resilient to the quantum noise on NISQ devices~\cite{caro2022generalization, schuld2019quantum}, highlighting the advantages of deploying VQC for implementing QML in many real-world applications. 

In addition to quantum computing for the machine learning paradigm, we are exploring how classical machine learning can contribute to developing QML. Given the current limitations of quantum computers, hybrid quantum-classical neural networks, combining classical and quantum components, are commonly used in QML to use available resources best. These hybrid architectures harness the speed advantages of quantum computing for specific tasks while relying on classical computing for operations it performs more effectively. This paper also focuses on enhancing QML's capabilities through classical machine learning techniques, aiming to improve its ability to represent data, generalize to new examples, and expand its applicability to real-world challenges.

In the discussion session, we offer fresh perspectives beyond the existing QML paradigms and explore how to harness the potential of cutting-edge generative AI and emerging quantum computing technologies in real-world industry use cases.

\section{Quantum Computing for Machine Learning}

Quantum computers with hundreds of logical qubits could significantly advance machine learning by offering dramatically faster performance than traditional methods. As illustrated in Table~\ref{tab1}, quantum algorithms can accelerate various machine learning tasks. For instance, neural networks and Boltzmann machines benefit from quadratic speedups, while PCA and SVM could see exponential improvements. This advantage stems from the unique capabilities of quantum circuits. While classical computers use simple gates like AND and OR, quantum logic gates are associated with unitary matrices and have much more representation capability than classical gates. This allows quantum computers to solve problems intractable to classical machines efficiently. 

\begin{table}[tpbh]\footnotesize
\vskip -0.1in
\center
\renewcommand{\arraystretch}{1.3}
\caption{Quantum computing speeds up classical machine learning algorithms~\cite{biamonte2017quantum}}
\begin{tabular}{|c||c|}
\hline
Methods     		& Speed-up	  \\
\hline
Quantum Boltzmann Machine	&	$\mathcal{O}(\sqrt{N})$		\\
\hline
Bayesian Inference  &   $\mathcal{O}(\sqrt{N})$	        \\
\hline
Quantum PCA		& $\mathcal{O}(\log N)$ 	 \\
\hline
Quantum SVM		& $\mathcal{O}(\log N)$		\\
\hline
Quantum Neural Network &  $\mathcal{O}(\sqrt{N})$		\\
\hline
Quantum Reinforcement Learning &  $\mathcal{O}(\sqrt{N})$ \\
\hline
\end{tabular}
\label{tab1}
\end{table}

One-qubit gates often include the Pauli-X, Y, and Z gates, while two-qubit gates typically involve the Controlled NOT (CNOT) and Controlled Z (CZ) gates. These quantum logic gates can be represented mathematically as follows:

\begin{equation*}
\rm X = \begin{bmatrix} 0 & 1 \\ 1 & 0 \end{bmatrix}, \hspace{3mm} Y = \begin{bmatrix} 0 & -i \\ i & 0 \end{bmatrix}, \hspace{3mm} Z = \begin{bmatrix} 1 & 0 \\ 0 & -1 \end{bmatrix}, 
\end{equation*}

\begin{equation*}
\rm CNOT = \begin{bmatrix} 1 & 0 & 0 & 0 \\ 0 & 1 & 0 & 0 \\ 0 & 0 & 0 & 1 \\ 0 & 0 & 1 & 0 \end{bmatrix},  \hspace{3mm} CZ = \begin{bmatrix} 1 & 0 & 0 & 0 \\ 0 & 1 & 0 & 0 \\ 0 & 0 & 1 & 0 \\ 0 & 0 & 0 & -1 \end{bmatrix}. 
\end{equation*}

According to the universal approximation theorem for quantum circuits, any continuous function can be approximated using a combination of one- and two-qubit gates. 
In the QML paradigm, we use Pauli rotation gates with adjustable angles to create a flexible quantum circuit structure. These angles, taken as the machine learning parameters, can be tuned to fit the specific data and task. 

\subsection{Variational Quantum Circuits}

\begin{figure}[t]
\vskip -0.2in
\centerline{\epsfig{figure=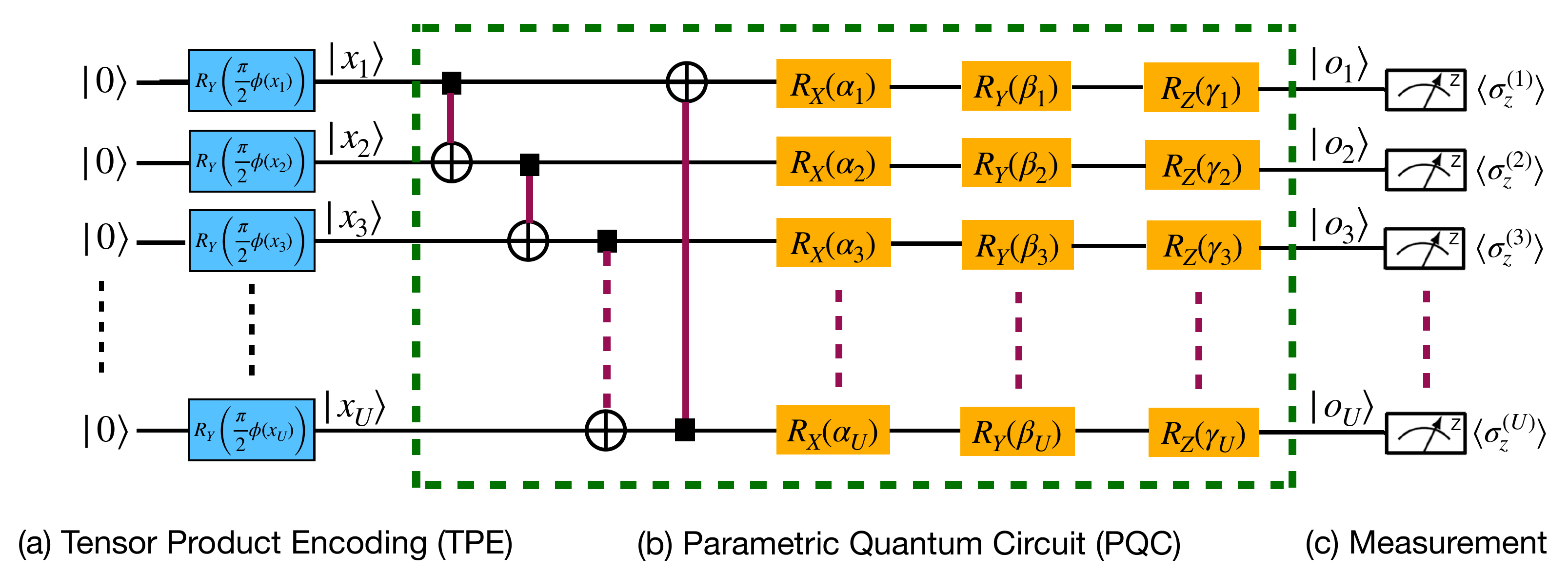, width=90mm}}
\caption{\textbf{Illustration of Variational Quantum Circuits}.}
\label{fig:vqc}
\vskip -0.2in
\end{figure}

The Pauli rotation gates are usually utilized to create variational quantum circuits (VQC). As shown in Figure~\ref{fig:vqc}, a typical VQC comprises angle encoding, parametric quantum circuits (PQC), and measurement. 

Given $U$ qubits, the angle encoding admits $U$ Pauli rotation gates $R_{Y}(\cdot)$ with fixed angles to constitute a tensor product encoding operation. Given a classical input vector $\textbf{x} = [x_{1}, x_{2}, ..., x_{U}]^{\top}$ into their corresponding quantum state $\vert \textbf{x} \rangle = [\vert x_{1} \rangle, \vert x_{2} \rangle, ..., \vert x_{U} \rangle]^{\top}$ through adopting a one-to-one mapping as: 

\begin{equation}
\label{eq:tpe}
\vert \textbf{x} \rangle = \left(\bigotimes_{i=1}^{U} R_{Y}(\frac{\pi}{2} \phi(x_{i})) \right) \vert 0 \rangle^{\otimes U},
\end{equation}
where $\phi(\cdot)$ refers to a non-linear function, e.g., $\phi(x_{i}) = \frac{1}{1 + \exp(-x_{i})}$ such that $\phi(x_{i})$ is restricted to the domain of $[0, 1]$ and the non-linearity is introduced in the feature space. 

In the PQC framework, we first implement quantum entanglement through a series of CNOT gates and then leverage Pauli rotation gates $R_{X}(\alpha_{i})$, $R_{Y}(\beta_{i})$ and $R_{Z}(\gamma_{i})$ with adjustable angles $\alpha_{i}$, $\beta_{i}$, $\gamma_{i}$ to construct a parametric circuit structure. The PQC model in the green dashed square is repeatedly copied to build a deep PQC architecture, which outputs $U$ quantum states $\vert o_{1} \rangle$, $\vert o_{2} \rangle$, ..., $\vert o_{U} \rangle$. The measurement transforms the quantum states $\vert o_{i} \rangle$ into the expected values $\langle \sigma_{i} \rangle = \langle o_{i} \vert \sigma_{z}^{(i)} \vert o_{i} \rangle$ associated with the observables of Pauli-Z matrices $\sigma_{z}^{(i)}$. We use an arithmetic average of $M$ times' quantum measurement to approximate $\langle \sigma_{i} \rangle$. 

\subsection{Quantum Reinforcement Learning}

\begin{figure}[h]
\vskip -0.15in
\centerline{\epsfig{figure=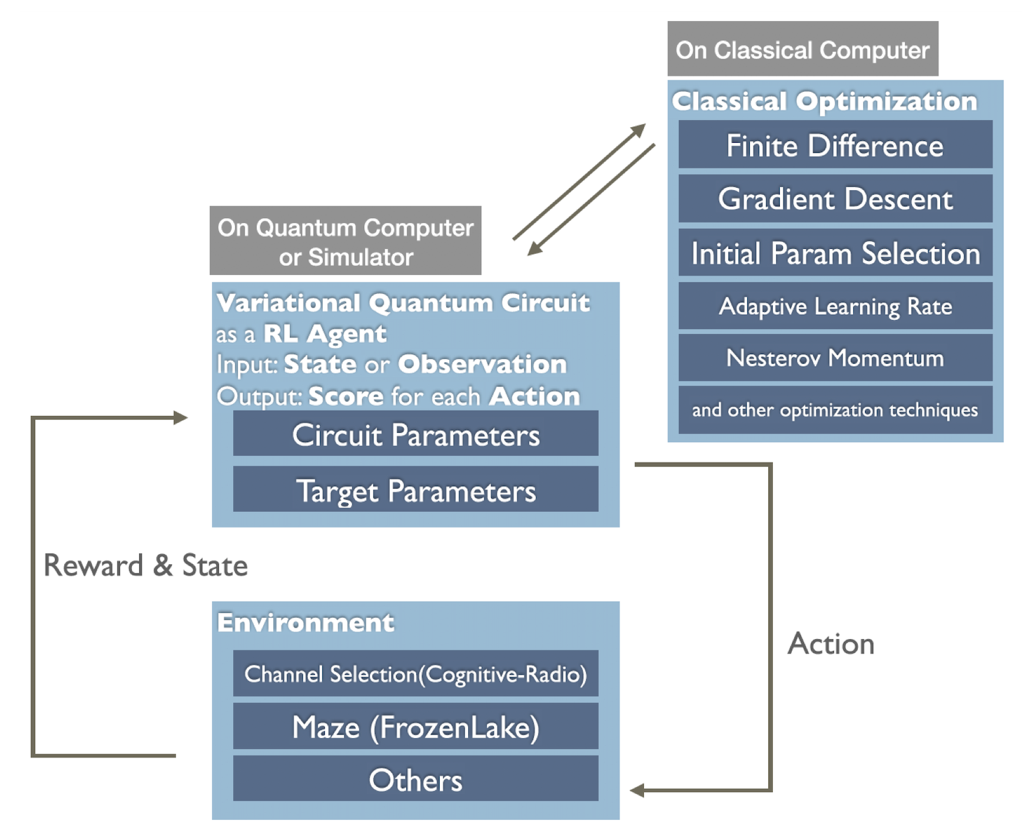, width=65mm}}
\caption{\textbf{Illustration of Quantum Reinforcement Learning}.}
\label{fig:qrl}
\vskip -0.15in
\end{figure}

A successful QML application based on the VQC structures in our work is quantum reinforcement learning (QRL)~\cite{chen2020variational,chen2022variational,chen2023QLSTM_RL,chen2023asynchronous,chen2023quantumDPER}. As illustrated in Figure~\ref{fig:qrl}, the VQC model, as the reinforcement learning agent, is processed on a quantum computer or quantum simulator. The classical computer selects the optimization techniques and controls quantum-classical interactions. 

Quantum Q-learning stands for the main approach to QRL. Quantum Q-learning, using a VQC as the agent, learns the best possible action-value function without following a specific policy. It starts with a random initialized $Q^{\pi}(s, a)$ for all states $s\in S$ and actions $a\in \mathcal{A}$, and it uses the VQC to represent $Q^{\pi}(s, a)$. The experience replay technique is employed to store past experiences as transition tuples $s_{t}$, $a_{t}$, $r_{t}$, $s_{t+1}$ in a CPU memory. After collecting sufficient experience information, the VQC agent randomly samples them to update the VQC to make it more stable. It also uses a separate target network to reduce the dependence on the current predictions. 

Quantum Q-learning is trained by minimizing an objective function like mean squared error (MSE), as shown in Eq. (\ref{eq:mse}). 
\begin{equation}
\label{eq:mse}
\mathcal{L}(\boldsymbol{\theta}) = \mathbb{E}\left[ r_{t} + \gamma \max\limits_{a'\in \mathcal{A}} Q(s_{t+1}, a'; \boldsymbol{\theta}^{-}) - Q(s_{t}, a_{t}; \boldsymbol{\theta})	\right]^{2}, 
\end{equation}
where $\gamma$ is a constant factor less than one used to discount future rewards, and $r_{t}$ is an immediate reward at time $t$ associated with the state $s_{t}$ and action $a_{t}$. Other loss functions can also be considered, e.g., Huber loss or mean absolute error (MAE)~\cite{qi2020mean}. In our work, QRL can solve environments with discrete observations, such as the Frozen Lake and Cognitive-Radio, where target network and experience replay are deployed. We also attempt to use quantum Q-learning for more sophisticated efforts in continuous observation spaces like Cart-Pole. 

\subsection{Quantum Convolutional Neural Network}

Quantum convolutional neural networks (QCNN)~\cite{cong2019quantum,chen2022quantumCNN} is a novel approach to neural networks that potentially leverage quantum computing principles to outperform classical convolutional neural networks (CNN) in many tasks. Instead of traditional convolution operations, QCNN employs quantum convolution, which involves applying the quantum gates to input qubits to extract features. This process can be more efficient than classical convolution, especially for high-dimensional data. Quantum convolution can extract more meaningful features from data due to quantum systems' inherent parallelism and entanglement. 

\begin{figure}[t]
\vskip -0.1in
\centerline{\epsfig{figure=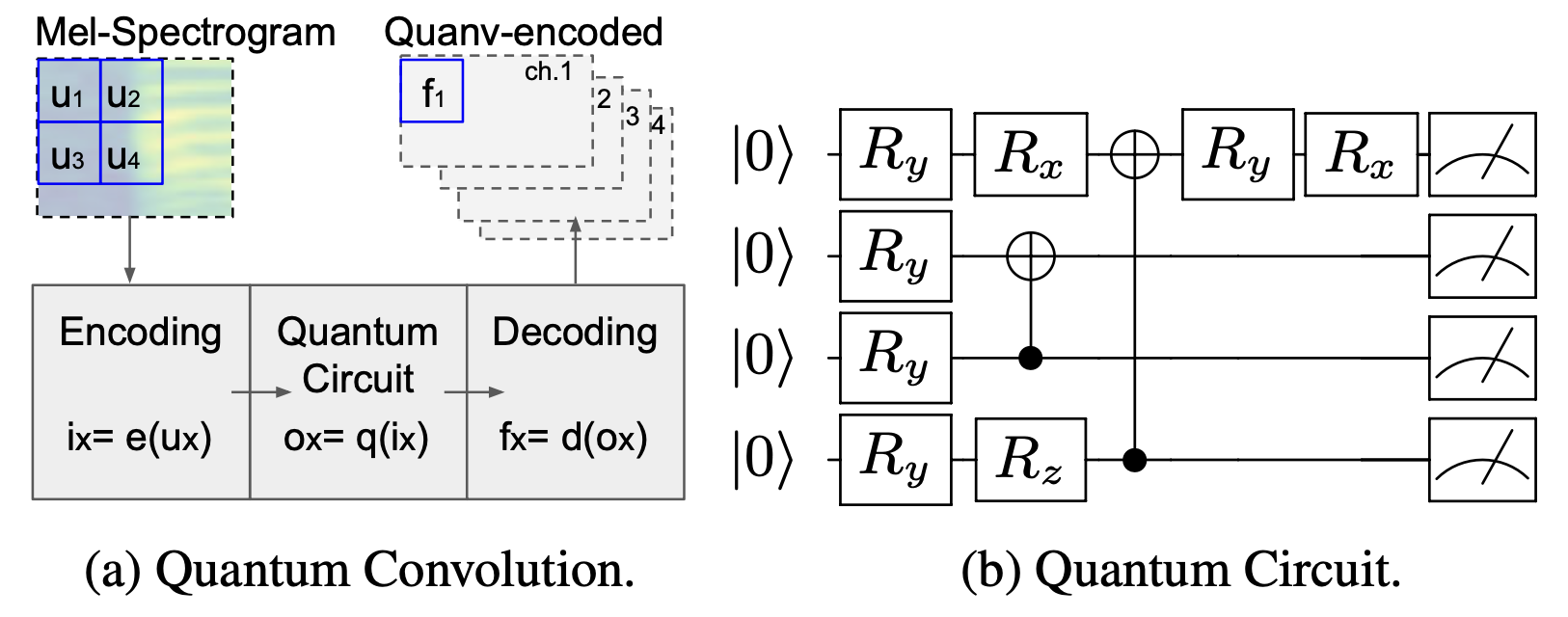, width=75mm}}
\caption{\textbf{An example of quantum circuit for convolution}.}
\label{fig:feature}
\vskip -0.1in
\end{figure}

\begin{figure}
\vskip -0.05in
\centerline{\epsfig{figure=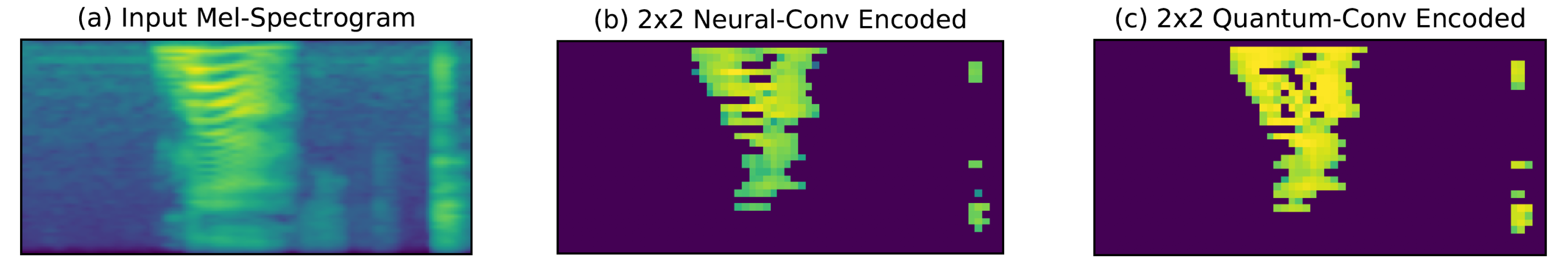, width=75mm}}
\caption{\textbf{A comparison of CNN and QCNN encoded speech signal features}.}
\label{fig:sig_res} 
\vskip -0.2in
\end{figure}

One of our real-world applications of QCNN is speech signal feature extraction with QCNN~\cite{yang2021decentralizing}. As shown in Figure~\ref{fig:feature}, we utilize the VQC structure to compose the quantum convolution that transforms the Mel-Spectrogram of speech signals into the corresponding QCNN encoded features. Figure~\ref{fig:sig_res} compares the signal signal features encoded by classical CNN and QCNN models. The QCNN exhibits a more discriminative representation of speech signals than the original Mel-Spectrogram and CNN encoded features, which results in even better speech recognition accuracy in our experiments of spoken language understanding. 

\section{Machine Learning Facilitates Quantum Machine Learning}

Our other research focuses on using machine learning techniques to facilitate QML development. Although the theoretical understanding of classical machine learning is far behind the rapid growth of the state-of-the-art empirical studies of generative AI, we can leverage the classical machine learning theory to scale up the QML for more complicated tasks. For one thing, we exploit a hybrid quantum-classical neural network to improve quantum models' representation and generalization powers. For another, we rely on the generative AI for quantum circuit architecture search. 

\subsection{Hybrid Quantum-Classical Neural Neworks}

\begin{figure}[t]
\vskip -0.1in
\centerline{\epsfig{figure=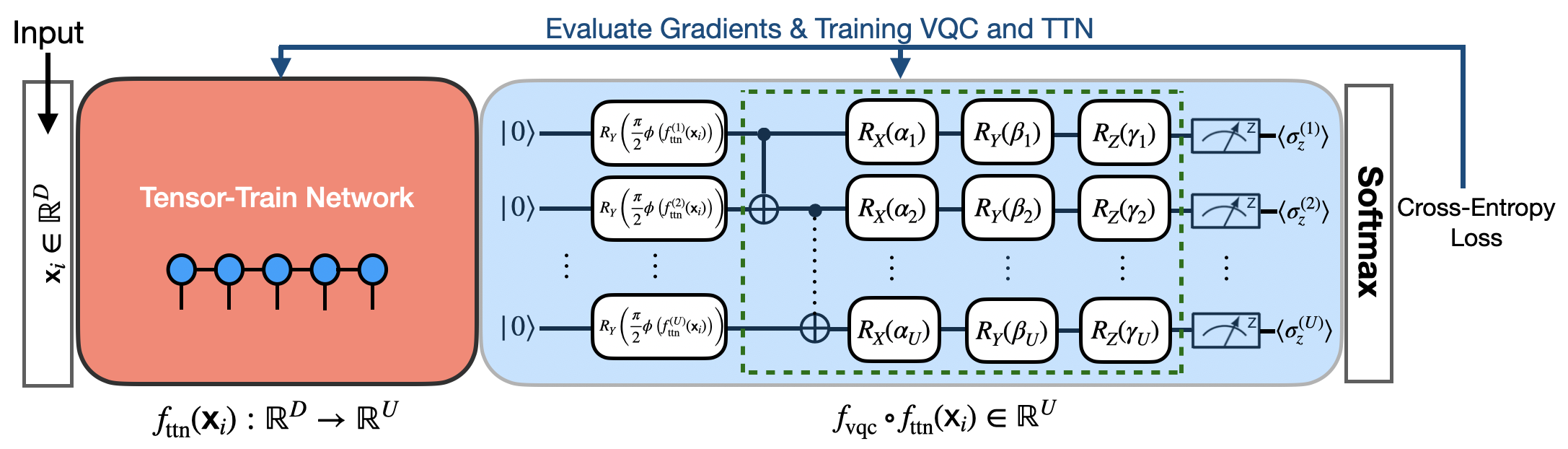, width=90mm}}
\caption{\textbf{An illustration of TTN-VQC structure}.}
\label{fig:ttn-vqc} 
\vskip -0.2in
\end{figure}

Hybrid quantum-classical neural networks combine the power of classical neural networks with the potential advantages of quantum components. These networks aim to harness the strengths of both paradigms to tackle complex problems that are challenging for classical or quantum systems alone~\cite{chen2021end}. 

Our previous work proposed an end-to-end quantum learning paradigm, TTN-VQC~\cite{qi2023qtn}, integrating a tensor-train network (TTN)~\cite{oseledets2011tensor, qi2022exploiting, omar2022mitigating} with the VQC structure as shown in Figure~\ref{fig:ttn-vqc}. TTN is a classical simulation of quantum circuits and provides a powerful and efficient way to represent high-dimensional tensors. When TTN is used on top of the VQC structure, TTN implements input feature dimensionality reduction while improving VQC's representation power~\cite{qi2023theoretical}. In particular, given $U$ qubits and $M$ times' measurement, the approximation error associated with the representation power is upper bounded by $\mathcal{O}(\frac{1}{\sqrt{U}}) + \mathcal{O}(\frac{1}{\sqrt{M}})$. However, the upper bound still relies upon the number of qubits, so the approximation error cannot be reduced to a small scale as we can access a small number of qubits. 

We proposed a transfer learning-based approach to improve the VQC's representation power~\cite{qi2023pre, qi2022classical, yang2022bert}. In this method, the classical TTN model is pre-trained and then integrated into the VQC structure. The resulting Pre+TTN-VQC architecture benefits from the TTN's pre-existing knowledge, making it less reliant on qubit constraints. Our experimental results of handwritten digit classification and semiconductor quantum dots corroborate our theoretical analysis. 

\begin{figure}[h]
\vskip -0.15in
\centerline{\epsfig{figure=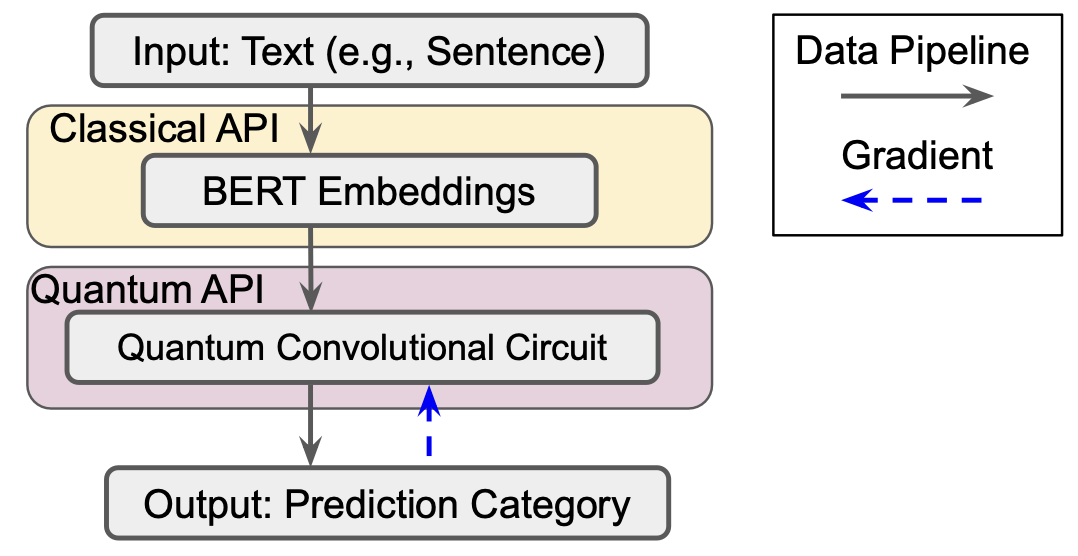, width=70mm}}
\caption{\textbf{Hybrid BERT-QCNN model for text classification}.}
\label{fig:bert} 
\end{figure}

As illustrated in Figure~\ref{fig:bert}, a concrete example of a hybrid quantum-classical neural network is the integration of BERT and QCNN for text classification~\cite{yang2022bert}. BERT~\cite{devlin2018bert}, a widely-used pre-trained language model, enhances the QCNN's quantum circuits' ability to represent text data. BERT's parameters remain unchanged in this hybrid architecture, while the VQC's parameters are fine-tuned for the specific task. Our classical simulations on CPU/GPU and real-world quantum experiments demonstrate that the BERT-QCNN model surpasses the performance of leading classical deep learning methods.

\subsection{Quantum Circuit Architecture Search}

Quantum circuit architecture search (QCAS) automatically designs optimal quantum circuits for specific tasks~\cite{du2022quantum}. It involves exploring a vast space of possible circuit configurations to find one that best balances performance, efficiency, and resource requirements. The search space of quantum circuit architectures includes factors like the number of qubits, gate types, and connectivity. We consider search algorithms to explore the search space and identify promising circuit architectures. 

\begin{figure}[t]
\vskip -0.1in
\centerline{\epsfig{figure=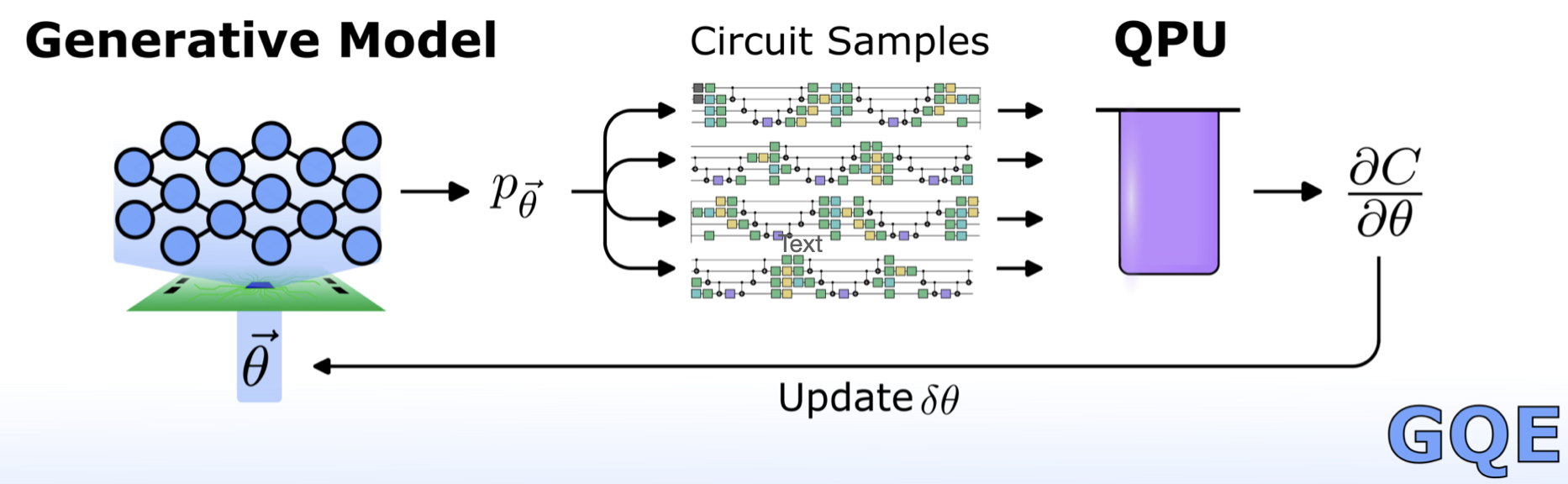, width=75mm}}
\caption{\textbf{Using generative models for quantum circuit architecture search}.}
\label{fig:genai} 
\vskip -0.1in
\end{figure}

Figure~\ref{fig:genai} showcases a generative model approach to designing quantum circuit architectures. Generative models, e.g., diffusion models, generate potential quantum circuit structures tailored to specific tasks~\cite{nakaji2024generative}. Rather than directly optimizing quantum circuits on a quantum processing unit (QPU), we refine the generative models to prioritize circuit designs that are more likely adequate for the given task. Generative models offer a potential solution to the optimization hurdles faced in VQC, allowing for exploring deeper quantum architectures.

Reinforcement learning (RL) provides another avenue for exploring quantum circuit architectures~\cite{kuo2021quantum,ye2021quantum,chen2023quantum,dai2024quantum}. An RL agent, often a neural network, generates and evaluates potential quantum circuit designs. The specific task or problem to be addressed serves as the environment for the RL agent. By learning from the RL rewards, e.g., accuracy or other relevant metrics, the RL agent refines its approach to find the most practical quantum circuit design. Further development combines the RL and QAS into a single framework using differentiable programming to optimize the VQC parameters and their architecture parameters simultaneously \cite{sun2023differentiable,chen2024differentiable}.

\begin{figure}[t]
\centerline{\epsfig{figure=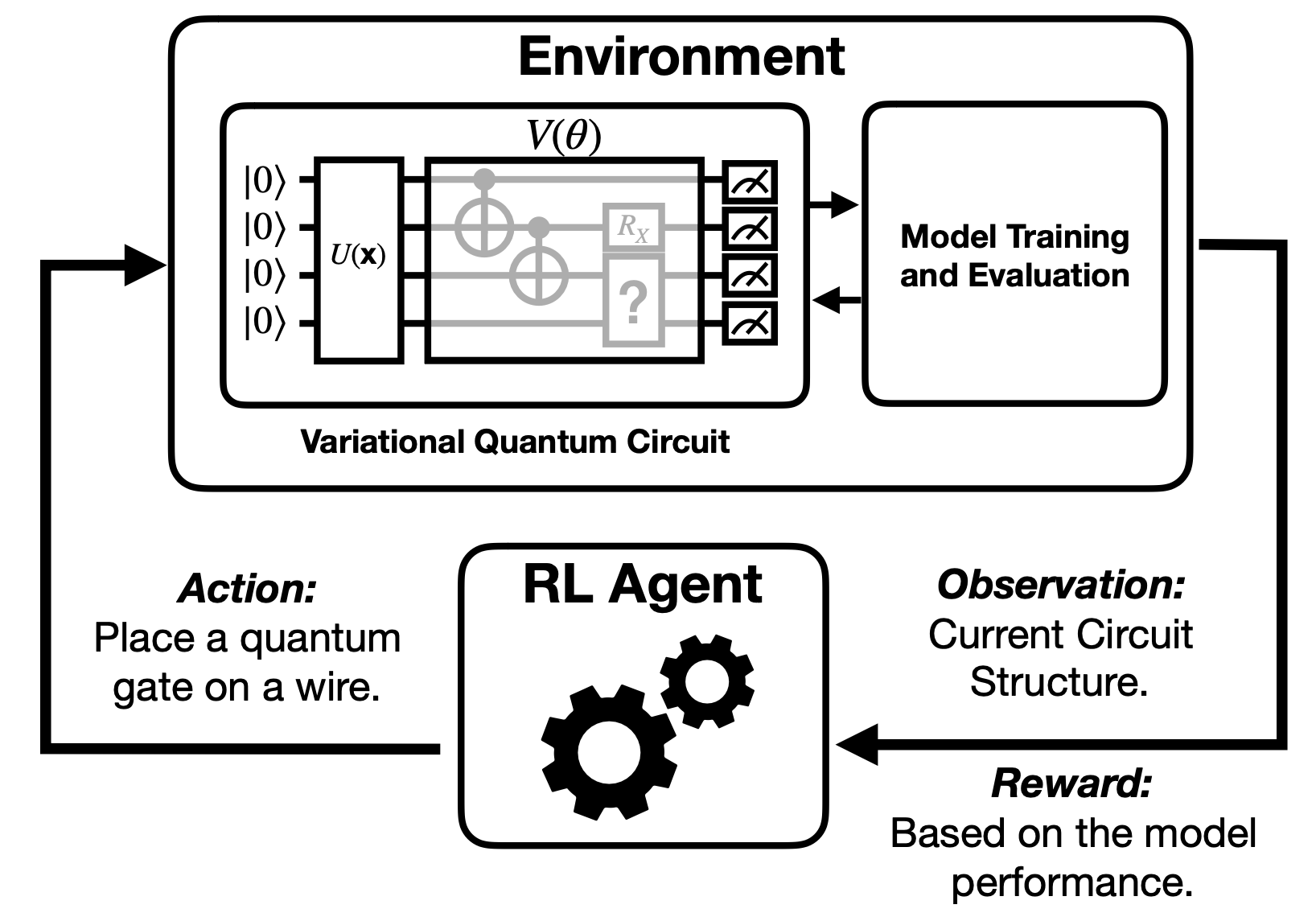, width=65mm}}
\caption{\textbf{Reinforcement learning for quantum circuit architecture search}.}
\label{fig:rl_arch} 
\vskip -0.2in
\end{figure}

\section{Discussions}

The intersection of machine learning and quantum computing is a rapidly evolving field with significant potential. We have explored novel QML approaches using quantum circuits to expedite specific machine-learning tasks. Besides, we applied classical machine learning principles to design innovative quantum algorithms, including hybrid quantum-classical neural networks and quantum circuit optimization techniques. During the NISQ era, classical machine learning methods, like pre-trained generative models and tensor networks, can enhance the performance of QML models, such as VQC, used in our work. However, our research primarily relies on classical simulations, assuming the existence of quantum logic qubits. For more practical use cases of QML, our future study of QML should consider the QML algorithms on realistic quantum computers. 

Quantum noise, a significant challenge in quantum computing, arises from interactions between qubits and their environments. Machine learning offers promising solutions to mitigate these noise-induced errors~\cite{cerezo2022challenges}. In particular, we mainly consider the critical approaches as follows: \\
1. \textbf{Quantum Error Correction Code (QEC) Optimization} 
	\begin{itemize}
	\item  \textit{Neural Networks}: Train neural networks to optimize QEC parameters, such as syndrome decoding rules, to minimize error rates. 
	\item  \textit{Reinforcement Learning}: Use RL agents to learn optimal error correction strategies, considering the quantum hardware's specific noise characteristics. 
	\end{itemize}
2. \textbf{Noise Characterization and Modeling} 
	\begin{itemize}
	\item  \textit{Generative Models}: Employ generative models like variational autoencoders or generative adversarial networks to learn the underlying distribution of noise patterns. 
	\item  \textit{Anomaly Detection}: Using machine learning techniques to identify anomalous noise events that deviate from expected patterns, enabling targeted mitigation strategies. 
	\end{itemize}
3. \textbf{Dynamic Noise Mitigation} 
	\begin{itemize}
	\item  \textit{Adaptive QEC}: Implement adaptive QEC schemes that adjust the error correction strategy based on real-time noise measurements. 
	\item  \textit{Machine Learning Based Gate Optimization}: Using machine learning to optimize gate sequences to minimize the impact of noise dynamically. 
	\end{itemize}
4. \textbf{Quantum Circuit Compilation}
	\begin{itemize}
	\item \textit{Noise-Aware Compilation}: Developing machine learning algorithms that compile quantum circuits to minimize noise-induced errors while preserving circuit functionality. 
	\item \textit{Error-Aware Gate Synthesis}: Using machine learning to synthesize quantum gates that are more robust to specific noise sources. 
	\end{itemize}

\bibliographystyle{IEEEbib}
\bibliography{sn-bibliography}

\begin{thebibliography}{10}

\bibitem{deng2018deep}
Li~Deng,
\newblock {\em {Deep Learning in Natural Language Processing}},
\newblock Springer, 2018.

\bibitem{szeliski2022computer}
Richard Szeliski,
\newblock {\em {Computer Vision: Algorithms and Applications}},
\newblock Springer Nature, 2022.

\bibitem{smalley2017ai}
Eric Smalley,
\newblock ``{AI-Powered Drug Discovery Captures Pharma Interest},''
\newblock {\em Nature Biotechnology}, vol. 35, no. 7, pp. 604--606, 2017.

\bibitem{nielsen2010quantum}
Michael~A Nielsen and Isaac~L Chuang,
\newblock {\em {Quantum Computation and Quantum Information}},
\newblock Cambridge University Press, 2010.

\bibitem{harper2019fault}
Robin Harper and Steven~T Flammia,
\newblock ``{Fault-Tolerant Logical Gates in The IBM Quantum Experience},''
\newblock {\em Physical review letters}, vol. 122, no. 8, pp. 080504, 2019.

\bibitem{kim2023cuda}
Jin-Sung Kim, Alex McCaskey, Bettina Heim, Manish Modani, Sam Stanwyck, and
  Timothy Costa,
\newblock ``{CUDA Quantum: The Platform for Integrated Quantum-Classical
  Computing},''
\newblock in {\em 2023 60th ACM/IEEE Design Automation Conference}. IEEE, 2023,
  pp. 1--4.

\bibitem{biamonte2017quantum}
Jacob Biamonte, Peter Wittek, Nicola Pancotti, Patrick Rebentrost, Nathan
  Wiebe, and Seth Lloyd,
\newblock ``{Quantum Machine Learning},''
\newblock {\em Nature}, vol. 549, no. 7671, pp. 195--202, 2017.

\bibitem{power_data}
Hsin-Yuan Huang et~al.,
\newblock ``{Power of Data in Quantum Machine Learning},''
\newblock {\em Nature Communications}, vol. 12, no. 1, pp. 1--9, 2021.

\bibitem{schuld2015introduction}
Maria Schuld, Ilya Sinayskiy, and Francesco Petruccione,
\newblock ``{An Introduction to Quantum Machine Learning},''
\newblock {\em Contemporary Physics}, vol. 56, no. 2, pp. 172--185, 2015.

\bibitem{liu2021rigorous}
Yunchao Liu, Srinivasan Arunachalam, and Kristan Temme,
\newblock ``{A Rigorous and Robust Quantum Speed-up in Supervised Machine
  Learning},''
\newblock {\em Nature Physics}, vol. 17, no. 9, pp. 1013--1017, 2021.

\bibitem{liu2024towards}
Junyu Liu et~al.,
\newblock ``{Towards Provably Efficient Quantum Algorithms for Large-scale
  Machine-Learning Models},''
\newblock {\em Nature Communications}, vol. 15, no. 1, pp. 434, 2024.

\bibitem{Preskill2018quantumcomputingin}
John Preskill,
\newblock ``{Quantum Computing in the NISQ Era and Beyond},''
\newblock {\em Quantum}, vol. 2, pp. 79, August 2018.

\bibitem{cerezo2022challenges}
M~Cerezo, Guillaume Verdon, Hsin-Yuan Huang, Lukasz Cincio, and Patrick~J
  Coles,
\newblock ``{Challenges and Opportunities in Quantum Machine Learning},''
\newblock {\em Nature Computational Science}, vol. 2, no. 9, pp. 567--576,
  2022.

\bibitem{cerezo2021variational}
Marco Cerezo et~al.,
\newblock ``{Variational Quantum Algorithms},''
\newblock {\em Nature Reviews Physics}, vol. 3, no. 9, pp. 625--644, 2021.

\bibitem{cong2019quantum}
Iris Cong, Soonwon Choi, and Mikhail~D Lukin,
\newblock ``{Quantum Convolutional Neural Networks},''
\newblock {\em Nature Physics}, vol. 15, no. 12, pp. 1273--1278, 2019.

\bibitem{verdon2019quantum}
Guillaume Verdon, Trevor McCourt, Enxhell Luzhnica, Vikash Singh, Stefan
  Leichenauer, and Jack Hidary,
\newblock ``{Quantum Graph Neural Networks},''
\newblock {\em arXiv preprint arXiv:1909.12264}, 2019.

\bibitem{caro2022generalization}
Matthias~C Caro et~al.,
\newblock ``{Generalization In Quantum Machine Learning From Few Training
  Data},''
\newblock {\em Nature Communications}, vol. 13, no. 1, pp. 4919, 2022.

\bibitem{schuld2019quantum}
Maria Schuld and Nathan Killoran,
\newblock ``{Quantum Machine Learning in Feature Hilbert Spaces},''
\newblock {\em Physical Review Letters}, vol. 122, no. 4, pp. 040504, 2019.

\bibitem{chen2020variational}
Samuel Yen-Chi Chen et~al.,
\newblock ``{Variational Quantum Circuits for Deep Reinforcement Learning},''
\newblock {\em IEEE Access}, vol. 8, pp. 141007--141024, 2020.

\bibitem{chen2022variational}
Samuel Yen-Chi Chen, Chih-Min Huang, Chia-Wei Hsing, Hsi-Sheng Goan, and
  Ying-Jer Kao,
\newblock ``{Variational quantum reinforcement learning via evolutionary
  optimization},''
\newblock {\em Machine Learning: Science and Technology}, vol. 3, no. 1, pp.
  015025, 2022.

\bibitem{chen2023QLSTM_RL}
Samuel Yen-Chi Chen,
\newblock ``{Quantum deep recurrent reinforcement learning},''
\newblock in {\em ICASSP 2023-2023 IEEE International Conference on Acoustics,
  Speech and Signal Processing (ICASSP)}. IEEE, 2023, pp. 1--5.

\bibitem{chen2023asynchronous}
Samuel Yen-Chi Chen,
\newblock ``{Asynchronous training of quantum reinforcement learning},''
\newblock {\em Procedia Computer Science}, vol. 222, pp. 321--330, 2023.

\bibitem{chen2023quantumDPER}
Samuel Yen-Chi Chen,
\newblock ``{Quantum deep Q-learning with distributed prioritized experience
  replay},''
\newblock in {\em 2023 IEEE International Conference on Quantum Computing and
  Engineering}. IEEE, 2023, vol.~2, pp. 31--35.

\bibitem{qi2020mean}
Jun Qi, Jun Du, Sabato~Marco Siniscalchi, Xiaoli Ma, and Chin-Hui Lee,
\newblock ``{On Mean Absolute Error for Deep Neural Network Based
  Vector-to-Vector Regression},''
\newblock {\em IEEE Signal Processing Letters}, vol. 27, pp. 1485--1489, 2020.

\bibitem{chen2022quantumCNN}
Samuel Yen-Chi Chen, Tzu-Chieh Wei, Chao Zhang, Haiwang Yu, and Shinjae Yoo,
\newblock ``{Quantum convolutional neural networks for high energy physics data
  analysis},''
\newblock {\em Physical Review Research}, vol. 4, no. 1, pp. 013231, 2022.

\bibitem{yang2021decentralizing}
Chao-Han~Huck Yang, Jun Qi, Samuel Yen-Chi Chen, Pin-Yu Chen, Sabato~Marco
  Siniscalchi, Xiaoli Ma, and Chin-Hui Lee,
\newblock ``Decentralizing feature extraction with quantum convolutional neural
  network for automatic speech recognition,''
\newblock in {\em ICASSP 2021-2021 IEEE International Conference on Acoustics,
  Speech and Signal Processing (ICASSP)}. IEEE, 2021, pp. 6523--6527.

\bibitem{chen2021end}
Samuel Yen-Chi Chen, Chih-Min Huang, Chia-Wei Hsing, and Ying-Jer Kao,
\newblock ``{An End-to-End Trainable Hybrid Classical-Quantum Classifier},''
\newblock {\em Machine Learning: Science and Technology}, vol. 2, no. 4, pp.
  045021, 2021.

\bibitem{qi2023qtn}
Jun Qi, Chao-Han Yang, and Pin-Yu Chen,
\newblock ``{QTN-VQC: An End-to-End Learning Framework for Quantum Neural
  Networks},''
\newblock {\em Physica Scripta}, vol. 99, no. 1, pp. 015111, 2023.

\bibitem{oseledets2011tensor}
Ivan~V Oseledets,
\newblock ``{Tensor-Train Decomposition},''
\newblock {\em SIAM Journal on Scientific Computing}, vol. 33, no. 5, pp.
  2295--2317, 2011.

\bibitem{qi2022exploiting}
Jun Qi, Chao-Han~Huck Yang, Pin-Yu Chen, and Javier Tejedor,
\newblock ``{Exploiting Low-Rank Tensor-Train Deep Neural Networks Based on
  Riemannian Gradient Descent With Illustrations of Speech Processing},''
\newblock {\em arXiv preprint arXiv:2203.06031}, 2022.

\bibitem{omar2022mitigating}
Muhammad~Shahmeer Omar, Jun Qi, and Xiaoli Ma,
\newblock ``{Mitigating Clipping Distortion in Multicarrier Transmissions Using
  Tensor-Train Deep Neural Networks},''
\newblock {\em IEEE Transactions on Wireless Communications}, vol. 22, no. 3,
  pp. 2127--2138, 2022.

\bibitem{qi2023theoretical}
Jun Qi, Chao-Han~Huck Yang, Pin-Yu Chen, and Min-Hsiu Hsieh,
\newblock ``{Theoretical Error Performance Analysis for Variational Quantum
  Circuit Based Functional Regression},''
\newblock {\em npj Quantum Information}, vol. 9, no. 1, pp. 4, 2023.

\bibitem{qi2023pre}
Jun Qi, Chao-Han~Huck Yang, Pin-Yu Chen, and Min-Hsiu Hsieh,
\newblock ``{Pre-training Tensor-Train Networks Facilitates Machine Learning
  with Variational Quantum Circuits},''
\newblock {\em arXiv preprint arXiv:2306.03741}, 2023.

\bibitem{qi2022classical}
Jun Qi and Javier Tejedor,
\newblock ``{Classical-to-Quantum Transfer Learning for Spoken Command
  Recognition Based on Quantum Neural Networks},''
\newblock in {\em IEEE International Conference on Acoustics, Speech and Signal
  Processing}, 2022, pp. 8627--8631.

\bibitem{yang2022bert}
Chao-Han~Huck Yang, Jun Qi, Samuel Yen-Chi Chen, Yu~Tsao, and Pin-Yu Chen,
\newblock ``{When BERT Meets Quantum Temporal Convolution Learning for Text
  Classification in Heterogeneous Computing},''
\newblock in {\em IEEE International Conference on Acoustics, Speech and Signal
  Processing}, 2022.

\bibitem{devlin2018bert}
Jacob Devlin, Ming-Wei Chang, Kenton Lee, and Kristina Toutanova,
\newblock ``{BERT: Pre-training of Deep Bidirectional Transformers for Language
  Understanding},''
\newblock in {\em Conference of the North American Chapter of the Association
  for Computational Linguistics: Human Language Technologies}, 2019, pp.
  4171--4186.

\bibitem{du2022quantum}
Yuxuan Du, Tao Huang, Shan You, Min-Hsiu Hsieh, and Dacheng Tao,
\newblock ``{Quantum Circuit Architecture Search for Variational Quantum
  Algorithms},''
\newblock {\em npj Quantum Information}, vol. 8, no. 1, pp. 62, 2022.

\bibitem{nakaji2024generative}
Kouhei Nakaji et~al.,
\newblock ``{The Generative Quantum Eigensolver (GQE) and Its Application for
  Ground State Search},''
\newblock {\em arXiv preprint arXiv:2401.09253}, 2024.

\bibitem{kuo2021quantum}
En-Jui Kuo, Yao-Lung~L Fang, and Samuel Yen-Chi Chen,
\newblock ``{Quantum architecture search via deep reinforcement learning},''
\newblock {\em arXiv preprint arXiv:2104.07715}, 2021.

\bibitem{ye2021quantum}
Esther Ye and Samuel Yen-Chi Chen,
\newblock ``{Quantum architecture search via continual reinforcement
  learning},''
\newblock {\em arXiv preprint arXiv:2112.05779}, 2021.

\bibitem{chen2023quantum}
Samuel Yen-Chi Chen,
\newblock ``{Quantum Reinforcement Learning for Quantum Architecture Search},''
\newblock in {\em International Workshop on Quantum Classical Cooperative},
  2023, pp. 17--20.

\bibitem{dai2024quantum}
Xin Dai, Tzu-Chieh Wei, Shinjae Yoo, and Samuel Yen-Chi Chen,
\newblock ``{Quantum Machine Learning Architecture Search via Deep
  Reinforcement Learning},''
\newblock {\em arXiv preprint arXiv:2407.20147}, 2024.

\bibitem{sun2023differentiable}
Yize Sun, Yunpu Ma, and Volker Tresp,
\newblock ``{Differentiable quantum architecture search for quantum
  reinforcement learning},''
\newblock in {\em 2023 IEEE International Conference on Quantum Computing and
  Engineering}. IEEE, 2023, vol.~2, pp. 15--19.

\bibitem{chen2024differentiable}
Samuel Yen-Chi Chen,
\newblock ``{Differentiable Quantum Architecture Search in Asynchronous Quantum
  Reinforcement Learning},''
\newblock {\em arXiv preprint arXiv:2407.18202}, 2024.

\end{thebibliography}

\end{document}